# Banking Stability System: Does it Matter if the Rate of Return is Fixed or Stochastic?

Hassan Belkacem Ghassan

Umm Al-Qura University, Department of Economics

Email: hbghassan@yahoo.com    hghassan@uqu.edu.sa

**Abstract**
The purpose is to compare the perfect Stochastic Return (SR) model like Islamic banks to the Fixed Return (FR) model as in conventional banks by measuring up their impacts at the macroeconomic level. We prove that if the optimal choice of share's investor in SR model $\alpha^*$ realizes the indifference of the financial institution toward SR and FR models, there exists $\alpha < \alpha^*$ such that the banks strictly prefers the SR model. Also, there exists $\alpha$, $\gamma$ and $\lambda$ verifying the conditions of α-sharing such that each party in economy can be better under the SR model and the economic welfare could be improved in a Pareto-efficient way.

# 1. Introduction

The purpose of this paper is to compare the stochastic return model i.e. applied by Islamic banks to the fixed return model i.e. followed by traditional banks. The comparison will be done by measuring up the impacts of each model at the macroeconomic level. The commercial financing model stipulates a fixed payoff as the retribution of financial capital use even if this latter does not generate a profit for the investor. In contrast, the Islamic financial contract must specify only the sharing of the return from the project between the investor and the financier.

The basic hypotheses are specified in order to prove which one of two models is more efficient and could improve social welfare (Khan, 1989). There are two forms of finance are available to investors in different economic activities, Stochastic Return Model (SR) and Fixed Return Model (FR). The Stochastic Return is based directly on the investment profitability, while Fixed Return is independent of the investment outcome. Nevertheless, historically the interest rate charged by banks has been considerably lower than the unobservable profit rate. Several reasons may be adduced in support of each hypothesis.

The hypotheses are related to: firstly supply and demand of financial funds in the financial market, secondly sharing information system and finally the attitudes to risks mainly for the macro-investor (Siddiqi, 2004). The monetary base, generated initially by a central bank, constitutes the supply of available financial funds in the economy which can be either used as credit or as corporation to productive sectors. The decisions depend upon the disposable contracts in the financial market. Two principal agents are concerned directly to make such decisions, banks or financial institutions as manager of finance flows, and the investors as manager of real flows.

The paper is organized as following, in second section economic, financial hypothesis and technical are presented. Third section displays the risk aversion under SR and FR models. Fourth section exhibits the determination of a best choice, fifth reveals the Pareto-efficiency, and last section concludes by the main findings.

# 2. Hypotheses
## 2.1 Economic and Financial Hypothesis

In fact, the aggregate saving is deduced from saving decisions which are based strictly on the level of income. So theoretically the aggregate saving of households does not depend on the interest rate even if they have deposits in banks. These financial resources are mobilized costly by banks in their credit system. At this stage, it is easy to argue that the saving decision and the investing decision are two distinct operations. A higher remuneration should be attached to the active one (Balestra and Baranzini, 1971).

In light of the previous definitions and principles it is possible to spell out the following hypotheses required for the comparison of SR's model to FR's model. The Financial market is unambiguous (H1), this assumption explains that in regular economic conditions, the investment projects are not correlated. When the business cycle is in the downswing, i.e. during a recession banks expect profit to decline, the losses would spread throughout the credit market. In this context, the FR's model leads to a series of bankruptcies of banks, whereas the SR's model constitutes an advantage because the two contractual parties would strive conjointly to minimize lose and could reorient in other way the investment project.

The hypothesis of Separation (H2) requires that the performance of project is no correlated to financing decision. Formally the production function is independent of whether the funds are from debt (FR) or equity in the common sense (SR)[1]. At least this absence of correlation between technical production and financial mode of activities of firm could be justify since the marginal productivity of capital, ceteris-paribus, should be independent of whether the new investment is raised via debt or equity. The efforts of investor are identical across different financial contracts. This assumption avoids the latent problem of moral hazard.

The drawback of this hypothesis is that it rules out incentive effects associated with the ownership structure of the firm. This is true essentially for FR's model, because the SR's model implies joint ownership. Furthermore, considering the investor, the major reason for its preference of debt is the existence of a *moral hazard* problem in the credit market. The hypothesis of Risk aversion (H3) indicates that the investors are not free of risk aversion. For the SRs model, this hypothesis is realistic and could be explained by the widespread use of equity in the firm's capital structure. While the existence of credit market and stock market reveals the aspiration of firms to build up various contracts forms for shifting risk to creditors. In fact, the choice of contract form depends more on the attitude of the investors towards risk.

---

[1] Here the concept of equity does not mean the sharing system of exchange stock. It is based on the direct participation in the production corporate i.e. sharing system of production with optimal distribution of risk.

The hypothesis of Identical return likelihood (H4) [2] suggests that the investors and banks have identical and costless information. This hypothesis is reliable to the rational expectations on each side either investor-producer or *investor-bank*. It supposes also weak differential of information structure. The alternative hypothesis could show that debt will be preferred even if the investor seems to be more risk averse and he can't reach the upper tail of the return likelihood. In fact the investor can costlessly observe the performance of his project. He can provoke asymmetrical information in the credit market. Especially since the debt requires minimal information of the project. Thus, the availability of costless and minimal information could generate the problem of *moral hazard*.

The SRs model does not have any collateral requirement. But banks to avoid this problem require collateral. In this way the FRs model ejects potential investors. As such the SRs model overcomes this problem and there may be a lot more investors in the economy with more risky investment but higher expected return.

### 2.2 Technical Hypotheses

The first one is related to the Funds and returns of investment (H5). Considering $L$ as the total supply of investment funds:

$$L = Z_1 + Z_2 ;$$

$$Z_1 = \beta L \quad \& \quad Z_2 = (1-\beta)L \quad \text{with} \quad 0 < \beta < 1$$

where $Z_i$ is the amount of funds financing investment projects according to the $i$ model; $i = 1$ corresponds to SR model and $i = 2$ represents the FR model.

Let $Z_i R_i$ be the amount of profits made by investors in $i$ model where $R_i$ is the rate of return on investment, it is considered a stochastic variable and having real and positive values:

---

[2] Stiglitz and Weiss (1981) showed that with imperfect information, the firms can be credit rationing. They argue that asymmetry of information between banks and borrowers, leads to the problems of adverse selection and moral hazard. Given lack of information available to banks concerning the true risk level of each loan, they charge interest rates according to the average expected level of risk.

$$R_i \sim \text{i.i.d.}(\mu_R, \sigma_R^2) \quad 0 < R_i < 1$$

where $\mu_R = E(R_i)$ and $\sigma_{R_i}^2 = V(R_i)$. This distribution implies that:

$$E(Z_i R_i) = Z_i E(R_i)$$
$$V(Z_i R_i) = Z_i^2 V(R_i)$$

Formally, the FR model is the dominant financial system based on loans. It consists to allow a quantity of money (principal) to the investor on the condition that the principal plus a fixed interest (as a fixed percent return on principal) is payable on a fixed future time by a fixed amount of money noted $Z_2(1+D)$ where $D$ is the interest rate charged by bank. If the earnings fall below this fixed amount, then a lesser amount will be paid:

$$\text{if } R_2 \geq D : \text{the lender will receive } D$$
$$\text{if } R_2 < D : \text{the lender will receive the entire } R_2$$

In contrast, formally the SR model is a financial system based on profit sharing. In this model the financier would receive a share of the project's return. In case of loss, no party gets anything.

The second one concerns the Aggregate payoffs of financier and investor (H6). It is supposed that no collateral is required from investors in two models. In the FR model, the financial contract stipulates:

$$P_2 = Z_2 \min(R_2, D) \qquad Y_2 = Z_2 \max(R_2 - D, 0) \qquad 0 < D < 1$$

where $P_2$ is a semi-stochastic variable and $Y_2$ is a stochastic variable, they represent the aggregate payoffs for banks (lenders) and for investors (borrowers) respectively.

Whereas, in the SR model, the financial contract stipulates:

$$P_1 = (1-\alpha)Z_1 R_1 \qquad Y_1 = \alpha Z_1 R_1 \qquad 0 < \alpha < 1$$

where $P_1$ and $Y_1$ are stochastic aggregate payoffs for financiers (with a share $1-\alpha$) and investors respectively. After these specifications, the problem is to determine the best or preferred contract in the financial market i.e. the Pareto-Optimal contract[3].

---

[3] The allocation of resources in an economy is Pareto-optimal (efficient), if it is not possible to change the allocation of resources in such a way as to make some one better off without making others worse off.

## 3. Expectation and Risk Aversion
### 3.1 Expectation of Payoffs
How we can determine the financier's attitude toward the financial model as the total supply of funds $L$ is allocated to projects according to SR and FR models? We should calculate the expected payoff under these models. In the FR model, the expected payoff for the financier is giving by:

$$E(P_2) = E[Z_2 \min(R_2, D)] = Z_2 E[\min(R_2, D)] = (1-\beta)L\, E[\min(R_2, D)]$$
$$V(P_2) = V[Z_2 \min(R_2, D)] = Z_2^2 V[\min(R_2, D)] = (1-\beta)LZ_2\, V[\min(R_2, D)]$$

In the SR model, it is given by:

$$E(P_1) = E[(1-\alpha)Z_1 R_1] = (1-\alpha)Z_1 E(R_1) = (1-\alpha)\beta L\, E(R_1)$$
$$V(P_1) = V[(1-\alpha)Z_1 R_1] = (1-\alpha)^2 Z_1^2 V(R_1) = (1-\alpha)^2 \beta LZ_1\, V(R_1)$$

Since the financier is supposed to be risk averse (H3) like investor, it is important to examine his expected utility (assumed to be continuous) in order to identify his preference for any model. Also the bargaining financial process is assumed continuous. The financier will be indifferent between SR and FR models if he will receive the same expected payoff.

### 3.2 Risk Aversion under SR and FR Models
We have two ways to determine the best contract. It is possible for the investor to find $D$ and $\alpha$ so that: $D, \alpha$ exist such that $E(Y_1) = E(Y_2)$. It remains that at this point, the choice of the contract depends on the preference of the financier.

Also, it is possible to assume that the financier assumed is risk neutral to find $D$ and $\alpha$ so that: $D, \alpha$ exist such that $E(P_1) = E(P_2)$. At this point, the choice of the contract depends on the preference of the investor. We choose this approach by assuming that the financier is risk averse. Because the investor represents the demand side and he can choose the appropriate source of funds, either from SR or FR, to financing his productive investment's efforts.

**Proposition 3.1**: Let a fixed $D$, $0 < \alpha < 1$ and $\beta \geq \dfrac{1}{2}$: then it exists $\alpha^* \in \Re^+$ such that $E(P_2) = E(P_1)$. And for $\alpha < \alpha^*$: then $E(P_2) < E(P_1)$. (See Proof in Annex)

This result indicates that the bank may become indifference between the two models. In contrast, if a greater portion of investment funds are allocated to the FR model, the previous result shows that the financier prefers strictly the FR model. So the expected payoff in FR model is great than in SR model for all $\alpha$. The choice of model is based on its expected payoff, the bank strictly prefers the SR model which gives a higher expected payoff.

At this stage, the choice of the contract depends on the attitude of investor toward risk. If the investor is risk neutral, the result can be apprehended via a specific version of Miller-Modigliani (MM) theorem (Miller and Modigliani, 1961; Stiglitz, 1969; Miller, 1988). The celebrated MM theorem asserts that the value of the firm, in a given risk class, is independent of its capital structure i.e. irrelevance of the choice of financial model. This theorem does not suppose the risk aversion, and then it can't capture this attitude of an investor.

## 4. Determination of a Best Choice

Hypothesis 3 of risk aversion raises some difficulties in reaching the solution to the best choice. In order to precisely find the solution, the expected utility of the payoff will be important, mainly the point where this utility is identical under the two models.

Firstly, a risk averter i.e. investor would prefer a less risky income flow among flows having the same expected payoff. Then it remains important to determine the riskiness associated to payoffs in the two models. From FR model and SR model, we have respectively:

$$Y_2 = Z_2 \, max(R_2 - D, 0) \text{ and } Y_1 = Z_1(\alpha R_1) = \alpha Z_1 R_1$$

and at $\alpha^*$, it exists $D^* : E(Y_1) = E(Y_2)$

Consider $U$ the aggregate utility function of the investors. $U$ is supposed bounded and $U'' < 0$ (i.e. risk aversion). Since the expected payoff of the investors is identical in SR and FR models if $E[U(Y_1)] = E[U(Y_2)]$. Under the risk aversion, the SR model will be preferred so that:

$$E[U(Y_1)] > E[U(Y_2)]$$

because the investor in FR model is more inclined to risk and his expected payoff would be less.

This result must be proved and can be expressed in the following proposition. To capture the proprieties of $R$, it is important to define the $\alpha-\text{sharing}$ rule and the Mean Preserving Spread (MPS) in the next definitions (Rothschild and Stiglitz, 1970 and 1971):

**Definition of $\alpha-\text{sharing}$:** An $\alpha-\text{sharing}$ rule is a real function $S(R)$ such that (Merton, 1992; Levy, 2006):

(i) $\quad 0 < S(R) < R < 1$

(ii) $\quad E(S(R)) = \alpha E(R) = \alpha \mu_{S(R)}$ where $\alpha$ is the sharing rate for investors.

(iii) $\quad \sum_k S_k(R) = 1$ where $k$ stands for shareholder in risk investment.

**Definition of MPS:** A Mean-Preserving Spread of $\alpha-\text{sharing}$ is defined as follows:

$$S_Y(R_2) = S_Y(R_1) + Z_2$$

where $S_Y(R_2)$ is a mean-preserving spread of $S_Y(R_1)$ and $Z_2$ is a random variable that is statistically independent of $S_Y(R_1)$ such that $E(Z_i) = \mu_{Z_i} = 0$ and $V(Z_i) = \sigma_{Z_i}^2 > 0$.[4] The stochastic variable $Y$ stands for the aggregate payoffs for investors. The theoretical mean will be defined as:

$$E[S_Y(R_2)] = E[S_Y(R_1)] + E(Z_2) = E[S_Y(R_1)]$$

and the preserving spread is defined as follows:

$$V[S_Y(R_2)] = V[S_Y(R_1) + Z_2] = \sigma_{R_1}^2 + \sigma_{Z_2}^2 > \sigma_{R_1}^2$$

Intuitively, it is easy to expect that the risk averse associated with $S(R_1)$ implies the preference of $S_Y(R_1)$ to $S_Y(R_2)$. Indeed, the risk aversion shows that $S_Y(R_2)$ is riskier than $S_Y(R_1)$.

**Proposition 4.1:** Consider $S_Y(R_2) = \max(R_2 - D_Y, 0)$ and $S_P(R_2) = \min(R_2, D_Y)$ where $S_Y$ and $S_P$ verify the conditions (i) to (iii). Then, for any concave and bounded utility function $U$ and any $\alpha-\text{sharing}$ rule $S(R)$ particularly $S_Y(R_1)$, we have

$$E[U(S_P(R_2))] > E[U(S_Y(R_1))] > E[U(S_Y(R_2))].$$

---

[4] This strong hypothesis can be replaced by the uncorrelated random variables. And any function of $S_Y(R_i)$ is uncorrelated of any function of $Z_i$.

(See Proof in Annex)

We can prove separately the double inequality straightforwardly by using a MPS of $\alpha-\text{sharing}$ and the Taylor expansion at degree two. This result exhibits that any risk aversion agent would prefer the $\alpha-\text{sharing}$ of $S_P(R_2)$ to $\alpha-\text{sharing}$ of $S_Y(R_1)$. In consequence, $S_P(R_2)$ is preferred to $S_Y(R_1)$ which is preferred to $S_Y(R_2)$. The proposition also exhibits by transitivity that the $\alpha-\text{sharing}$-investor of $S_Y(R_1)$ would be preferred to the $\alpha-\text{sharing}$ of $S_Y(R_2)$. This result shows that in terms of $\alpha-\text{sharing}$ and of expected payoff the SR model seems to be better than the FR model.

## 5. Pareto Efficiency

This central concept is widely accepted standard for comparing economic outcomes. There are many efficient situations in which the active agent tends to avoid the inefficiency. Our proposition identifies inefficiency situations and it is important to designing policies and institutions that will promote efficiency and reduce inefficiency.

The financier in FR model in order to avoid many problems like moral hazard, adverse selection, bankruptcy could reduce Pareto-efficiently his weak risk aversion by improving the outcomes of investors in economy through the reallocation of payoffs. It is optimal that the financier immunes all investors (in macroeconomic view) and adopts the SR model. This must increase the correlation between the ownership and capital structure such that we have next Pareto-Optimal payoffs:

$$E[U(\tilde{S}_P(R_2))] \approx E[U(\tilde{S}_Y(R_1))] \approx E[U(\tilde{S}_Y(R_2))]$$

**Proposition 5.1:** Each party (investors and financiers) can be better in SR model and the economic welfare could be improved in a Pareto efficiency way. (See Proof in Annex)

## 6. Conclusion

The SR model seems to be Pareto-Optimal Contract and proves the inferiority of the FR contract i.e. debt model. But the credit market dominates throughout the world.

Furthermore, the Pareto-optimal solution could improve the payoff of financier in the FR model if the payoff of SR model has been increased and the additional outcomes would be distributed in half with the other investors operating in FR model.

These arrangements permit that these moves are likely to be Pareto-Optimal. The SR model has the spreading risk more evenly than the FR model, and considering the risk aversion of the investors, then the SR model dominates the FR model and the higher outcomes should be associated with the active one i.e. the SR investor. In addition, the SR model has far-reaching impacts for the stability of a financial system. So this result is very opportune for the financial policy purposes. The real macroeconomic investment might be higher in the SR model than in the FR model at least because it doesn't require imperative collateral and it could reward the riskier agent.

**Annex**

Proof proposition 3.1: Consider a continuous function $h(\alpha)$: $h(\alpha) = E(P_2) - E(P_1)$

$h(\alpha) = (1-\beta)LE[\min(R_2, D)] - (1-\alpha)\beta LE(R_1)$

$h(0) = L[(1-\beta)E[\min(R_2, D)] - \beta E(R_1)]$ and $h(1) = L[(1-\beta)E[\min(R_2, D)]] > 0$

If $\beta = \frac{1}{2}$: $h(0) = \frac{1}{2}L[E[\min(R_2, D) - E(R_1)]] \leq 0$, (by using intermediate value theorem). If $\beta \geq \frac{1}{2}$: $h(0) \leq 0$, but if $\beta < \frac{1}{2}$: $h(0) > 0$. As $h$ is continuous and $\beta \geq \frac{1}{2}$, it exists $\alpha^* \in \Re^+$: $h(0) \leq h(\alpha^*) \leq h(1)$ and $h(\alpha^*) = 0$. For $0 < \alpha < \alpha^*$: $h(\alpha) < 0 \Rightarrow E(P_1) > E(P_2)$ ∎

Proof proposition 4.1: Letting $S_Y(R_1)$ the random $\alpha$-sharing rate, and $U$ the utility function of $S(R)$ which is assumed with risk averse i.e. $U''(S(R)) < 0$. The expected utility of $S(R)$ is $E[U(S(R))]$.

$$U(S_Y(R_2)) = U(S_Y(R_1) + Z_2)$$
$$\approx U(S_Y(R_1)) + U'(S_Y(R_1))Z_2 + \frac{1}{2}U''(S_Y(R_1))Z_2^2$$

Then, the expected value gives:

$$E[U(S_Y(R_2))] = E[U(S_Y(R_1))] + E[U'(S_Y(R_1))Z_2] + \frac{1}{2}E[U''(S_Y(R_1))Z_2^2]$$
$$= E[U(S_Y(R_1))] + \frac{1}{2}E[U''(S_Y(R_1))]E(Z_2^2), \quad E(Z_i) = 0$$

$E[U(S_Y(R_1))] > E[U(S_Y(R_2))]$, $E[U''(S_Y(R_1))] < 0$ and $\sigma_{Z_2}^2 > 0$

In the same way, the second inequality shows that $S_Y(R_1)$ is riskier than $S_P(R_2)$. By using a MPS there exists $Z_1$ is a random variable such that $S_Y(R_1) = S_P(R_2) + Z_1$ which $S_P(R_2)$ is a random $\alpha$-sharing rate. The expected utility gives:

$$E[U(S_Y(R_1))] = E[U(S_P(R_2))] + E[U'(S_P(R_2))Z_1] + \frac{1}{2}E[U''(S_P(R_2))Z_1^2]$$

$$= E[U(S_P(R_2))] + \frac{1}{2}E[U''(S_P(R_2))]E(Z_1^2), \quad E(Z_i) = 0$$

$$E[U(S_P(R_2))] > E[U(S_Y(R_1))], \quad E[U''(S_P(R_2))] < 0 \text{ and } \sigma_{Z_1}^2 > 0. \quad \blacksquare$$

Proof proposition 5.1: From the proposition 1, for all $\alpha \in ]0, \alpha^*[ : h(\alpha) < 0$ and $E(P_1) > E(P_2)$ i.e. the financier would prefer SR model. $U$ is continuous and concave (i.e. risk aversion):

it exists $(1-\alpha) \in ]\alpha^*, 1[; \lambda \in ]0, \alpha^*[ : 1 - \alpha + \lambda = 1 - \alpha^*$ such that

$$E[U(S_P(R_2))] = E[U(S_P(R_1))] \Leftrightarrow E[U(\min(R_2, D_P))] = E[U(1-\alpha+\lambda)R_1]$$

This solution is Pareto-optimal. But it is possible to find another solution which can improve the payoff of the banks in FR model, and the lenders strictly prefer the SR model: it exists $(\alpha, \frac{\lambda}{2}) \in ]0, \alpha^*[$ such that $1 - \alpha + \frac{\lambda}{2} = 1 - \alpha^*$

$$E[U(\min((1+\tfrac{1}{2}\lambda)R_2, D_P))] = E[U(1-\alpha+\tfrac{1}{2}\lambda)R_1]$$

Also, it exists $(\alpha, \gamma) \in ]0, \alpha^*[ : \alpha + \gamma = \alpha^*$ such that

$$E[U(S_Y(R_2))] = E[U(S_Y(R_1))] \Leftrightarrow E[U(\max(R_2 - D_Y, 0))] = E[U(\alpha+\gamma)R_1]$$

It is possible to improve in Pareto-Optimal way the payoffs of investors in FR without changing the investor preference for the SR model, and the borrowers strictly prefer the SR model: it exists $(\alpha, \frac{\gamma}{2}) \in ]0, \alpha^*[$ such that $\alpha + \frac{\gamma}{2} = \alpha^*$

$$E[U(\max((1+\tfrac{1}{2}\gamma)R_2 - D_Y, 0))] = E[U(\alpha+\tfrac{1}{2}\gamma)R_1]. \quad \blacksquare$$